\documentclass[12pt,a4paper]{scrartcl}
\pdfoutput=1
\usepackage{amsmath,amssymb}
\usepackage{subfigure}

\usepackage{hyperref}
\usepackage{graphicx}
\usepackage{feynmf} 
\let\tmptitle\title\renewcommand{\title}[1]{\tmptitle{\LARGE #1}}
\let\tmpauthor\author\renewcommand{\author}[1]{\tmpauthor{\large #1}}
\let\tmpdate\date\renewcommand{\date}[1]{\tmpdate{\normalsize #1}}
\newcommand{\abstrct}[1]{\begin{abstract}\vspace{-2em}\small\noindent#1\end{abstract}}

\title{\Large
New Solution for Neutrino Masses and Leptogenesis in Adjoint $SU(5)$
}
\date{\today}
\author{
Kristjan Kannike$^{a,b}$\footnote{Email:
kristjan.kannike@sns.it} ~and Dmitry~V.~Zhuridov$^a$\footnote{Email: dmitry.zhuridov@sns.it}
\\ \normalsize\itshape
$^a$Scuola Normale Superiore and INFN, Piazza dei Cavalieri 7, 56126 Pisa, Italy 
\\ \normalsize\itshape
$^b$NICPB, Ravala 10, 10143 Tallinn, Estonia
}

\begin{document}

\maketitle

\abstrct{%
We investigate baryogenesis via leptogenesis and generation of neutrino masses and mixings through the Type I plus Type III seesaw plus an one-loop mechanism in the context of Renormalizable Adjoint $SU(5)$ theory. 
One light neutrino remains massless, because the contributions from three heavy Majorana fermions $\rho_0$, $\rho_3$ and $\rho_8$ to the neutrino mass matrix are not linearly independent. 
However none of these heavy fermions is decoupled from the generation of neutrino masses.
This opens a new range in parameter space for successful leptogenesis, in particular, allows for inverted hierarchy of the neutrino masses.
}

\begin{fmffile}{bkz}

\section{Introduction}

The observable baryon asymmetry of the Universe (BAU) and the nonzero neutrino masses~\cite{PDG2010} provide strong evidences of physics beyond the Standard Model (SM). 
The possibility of unification of the gauge interactions is one of the most intriguing guiding principles for building theories beyond the SM, in this case called the grand unified theories (GUT). Interestingly, one of the most economical realistic GUTs may explain both BAU and neutrino masses.

It is well known that $SU(5)$ is the minimal group that can provide for grand unification. The first and simplest GUT, proposed in \cite{Georgi_Glashow}, is based on this group. 
Its scalar sector contains only two representations of ${\bf 5}_H$ and ${\bf 24}_H$, while each generation of matter fields is embedded in two representations of $\bar{\bf5}$ and {\bf10}.
The $SU(5)$ symmetry is broken down to the SM one by the vacuum expectation value (VEV) of the scalar singlet in ${\bf 24}_H$, and the SM Higgs belongs to ${\bf 5}_H$.
In spite of its elegance, this model is ruled out since it can not: 
\begin{itemize}
 \item Unify the measured SM gauge couplings;
 \item Explain nonzero neutrino masses;
 \item Unify the SM Yukawa couplings of charged leptons and down quarks at high scale in the renormalizable level;
 \item Explain present experimental lower bound on the proton lifetime of order $10^{33}$~years.
\end{itemize} 
The first two of these problems were solved by adding a new fermionic {\bf 24} representation~\cite{0612029}. 
However a whole set of high dimensional operators was needed to address the third problem. 
The renormalizable generalization of the model of \cite{0612029} with the minimal number of scalar fields contains only one new scalar ${\bf 45}_H$ representation~\cite{0702287}, 
so that the electroweak symmetry is broken by both $\langle{\bf 5}_H\rangle$ and $\langle{\bf 45}_H\rangle$ VEVs, providing for realistic Yukawas. 
Ultimately, this nonsupersymmetric Renormalizable Adjoint $SU(5)$ theory contains scalars in ${\bf 5}_H$, ${\bf 24}_H$ and ${\bf 45}_H$, and matter in $\bar{\bf 5}_\alpha$, ${\bf 10}_\alpha$ ($\alpha=1,2,3$) and ${\bf 24}$.
(See Appendix~\ref{app:fields} for the details.)

It is well known~\cite{0702287} that the neutrino masses may be generated through Type I~\cite{seesawI} and Type III~\cite{seesawIII} seesaw mechanisms in the Renormalizable Adjoint $SU(5)$ model.
In addition, the BAU may be explained~\cite{LG} by generating the lepton asymmetry in the out-of-equilibrium decays of fermion triplets responsible for the Type III seesaw, 
and converting it to a baryon asymmetry by sphaleron transitions~\cite{KRS} in the usual baryogenesis~\cite{Sakharov} via leptogenesis (LG) scenario~\cite{Fukugita_Yanagida}.
In this paper we investigate a new range of the parameters where neutrino masses get a sizable contribution from the one-loop coloured seesaw besides the tree level type I and III seesaw mechanisms. 
However, since the loop term is proportional to a linear combination of type I and III seesaw Yukawas, the lightest neutrino still remains massless. With the new contribution to neutrino mass, 
the allowed range in the parameter space for successful LG is enlarged.

One of the interesting results of this model of the neutrino masses and LG in the framework of Adjoint $SU(5)$ theory is the upper bound on the proton lifetime of order $10^{35}$~years~\cite{0803.4156,0809.2106}, 
which can be tested in future. We remark that there is a parameter range in Adjoint $SU(5)$, which is hard to test at the proton decay experiments~\cite{Dorsner}. 
However within this range the observable neutrino masses and BAU can not be generated.

In the next section we investigate generation of the neutrino masses together with the unification of the couplings in the considered model. 
In section~\ref{section:nu_masses_exp} we discuss parametrization of the Yukawa couplings in terms of the neutrino experimental data.
In section~\ref{section:LG} we give formulae for LG. 
We discuss the numerical results in section~\ref{sec:numerics} and conclude in section~\ref{sec:summary}.

\section{Generation of neutrino masses}\label{section:nu_masses_generation}

The interactions relevant for the generation of masses of the adjoint fermions are given by
\begin{eqnarray}
 -{\mathcal{L}}_{m_{24}} &=& M \,{\rm Tr}\, {\bf 24}^2 + \lambda\, {\rm Tr}\,({\bf 24}^2{\bf 24}_H) + {\rm H.c.} \\
 &=& M\,24^i_{\ j}24^j_{\ i} + \lambda\,24^i_{\ j}24^j_{\ k}(24_H)^k_{\ i} + {\rm H.c.}, \nonumber
\end{eqnarray}
with $i,j,k=1,\dots,5$. 
The masses of the fermions responsible for the neutrino masses are given by~\cite{0803.4156}
\begin{equation}
\begin{split}
 M_{\rho_0}&=\left|M-\frac{\lambda M_{\rm GUT}}{\sqrt{50\pi \alpha_{\rm GUT}}}\right|, \;\,
 M_{\rho_3}=\left|M-\frac{3 \lambda M_{\rm GUT}}{\sqrt{50\pi \alpha_{\rm GUT}}}\right|, \;\,
 M_{\rho_8} =\left|M+\frac{2 \lambda M_{\rm GUT}}{\sqrt{50\pi \alpha_{\rm GUT}}}\right|,
 \end{split}
\end{equation}
where the scale of Grand Unification is  $M_{\text{GUT}}=v_{24}\sqrt{5\pi\alpha_{\text{GUT}}/3}$.
The regime of successful unification determines the approximate mass relations
\begin{eqnarray}\label{mass_relations}
 M_{\rho_3}\ll \hat m M_{\rho_3} = M_{\rho_8} \simeq \frac{5}{2}M_{\rho_0} 
\end{eqnarray} 
with $\hat m\gtrsim 10^2$.

For the mass spectrum of Adjoint SU(5), which is allowed by unification, proton decay and LG, see Fig.~\ref{fg:spectrum} 
with the description in section~\ref{sec:numerics}.

The Yukawa interactions needed to generate the neutrino masses are given by
\begin{eqnarray}
 -{\mathcal{L}}^{\text{Yukawa}} &=& c_\alpha{\bf\bar5_\alpha}{\bf24\ 5}_H + p_\alpha{\bf\bar5_\alpha}{\bf24\ 45}_H + {\rm H.c.} \nonumber \\
       &=& c_\alpha\bar5_{\alpha j} 24^j_{\ k}5^k_H + p_\alpha\bar5_{\alpha j} 24^k_{\ l}(45_H)^{jl}_k + {\rm H.c.},
\end{eqnarray}
which contains
\begin{eqnarray}
 -{\mathcal{L}}_\nu^{\text{Yukawa}}
       &=& c_\alpha\bar\ell_\alpha^c{\rm i}\sigma_2\rho_3H_1 + \frac{3c_\alpha}{2\sqrt{15}}\bar\ell_\alpha^c{\rm i}\sigma_2H_1\rho_0 \nonumber\\
       &-& 3p_\alpha\bar\ell_\alpha^c{\rm i}\sigma_2\rho_3H_2  + \frac{\sqrt{15}}{2}p_\alpha\bar\ell_\alpha^c{\rm i}\sigma_2H_2\rho_0 
       - \frac{p_\alpha}{\sqrt{2}}\bar\ell_\alpha^c{\rm i}\sigma_2\text{Tr}(S_8\rho_8) + {\rm H.c.},
\end{eqnarray}
where $\bar\psi^c=\psi^TC$ and the definitions used for the representations and multiplets are given in the Appendices~\ref{app:fields} and \ref{app:multiplets}.

\begin{figure}[h]
\centering
\subfigure[]{
\fmfframe(1,6)(0,-10){
\begin{fmfgraph*}(30,25)
\fmfleft{i1}
\fmfright{o1}
\fmftop{t1,t2,t3,t4}
\fmf{fermion,label=$\nu_{L\alpha}$,label.side=right}{i1,v1}
\fmf{fermion,tension=2}{v1,v3}
\fmf{fermion,tension=2}{v2,v3}
\fmfv{label=$\rho_3$,label.angle=-90}{v3}
\fmf{fermion,label=$\nu_{L\beta}$,label.side=left}{o1,v2}
\fmffreeze
\fmf{scalar}{t2,v1}
\fmf{scalar}{t3,v2}
\fmflabel{$\langle H\rangle$}{t2}
\fmflabel{$\langle H\rangle$}{t3}
\end{fmfgraph*}
} }
\subfigure[]{
\fmfframe(1,6)(0,-10){
\begin{fmfgraph*}(30,25)
\fmfleft{i1}
\fmfright{o1}
\fmftop{t1,t2,t3,t4}
\fmf{fermion,label=$\nu_{L\alpha}$,label.side=right}{i1,v1}
\fmf{fermion,tension=2}{v1,v3}
\fmf{fermion,tension=2}{v2,v3}
\fmfv{label=$\rho_0$,label.angle=-90}{v3}
\fmf{fermion,label=$\nu_{L\beta}$,label.side=left}{o1,v2}
\fmffreeze
\fmf{scalar}{t2,v1}
\fmf{scalar}{t3,v2}
\fmflabel{$\langle H\rangle$}{t2}
\fmflabel{$\langle H\rangle$}{t3}
\end{fmfgraph*}
} }
\subfigure[]{
\fmfframe(1,6)(0,-10){
\begin{fmfgraph*}(30,25)
\fmfleft{i1}
\fmfright{o1}
\fmftop{t1,t2,t3,t4}
\fmf{fermion,label=$\nu_{L\alpha}$,label.side=right}{i1,v1}
\fmf{fermion}{v1,v2}
\fmf{fermion}{v3,v2}
\fmf{fermion,label=$\nu_{L\beta}$,label.side=left}{o1,v3}
\fmfv{label=$\rho_8$,label.angle=-90}{v2}
\fmffreeze
\fmf{phantom,left=0.5,tension=0.5}{v1,v4,v3}
\fmf{scalar,label=$S_8$,label.side=right,right=0.5,tension=0.5}{v4,v1}
\fmf{scalar,label=$S_8$,label.side=left,left=0.5,tension=0.5}{v4,v3}
\fmf{scalar}{t2,v4}
\fmf{scalar}{t3,v4}
\fmflabel{$\langle H\rangle$}{t2}
\fmflabel{$\langle H\rangle$}{t3}
\end{fmfgraph*}
} }
\caption{The interactions that generate neutrino masses. The arrows in fermionic lines show flow of lepton number.}
\label{fig:1:loop:nu:masses}
\end{figure}
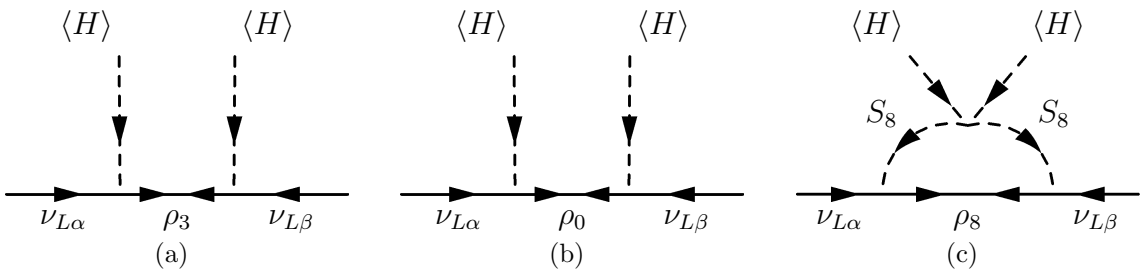

The scalar doublets $H_1$ and $H_2$ are related to the SM Higgs $H$ and the heavy scalar $H^\prime$ by the orthogonal transformation
\begin{eqnarray}
 \left(
\begin{array}{c}
H_1 \\
H_2 \\
\end{array}
\right) = \left(
\begin{array}{cc}
\cos\theta & -\sin\theta \\
\sin\theta & \cos\theta \\
\end{array}
\right) \left(
\begin{array}{c}
H \\
H^\prime \\
\end{array}
\right),
\end{eqnarray}
with~\cite{Dorsner_VEV}
\begin{equation}
 \cos\theta=\pm\frac{|v_5|}{v}, \quad \sin\theta=\pm2\sqrt{6}\frac{|v_{45}|}{v}.
\end{equation}
The VEVs of physical fields are $\langle H\rangle=v/\sqrt{2}\equiv v_0=174$~GeV and $\langle H^\prime\rangle=0$.
Hence the terms relevant for the neutrino masses generation at tree level, shown in Fig.~\ref{fig:1:loop:nu:masses}~$a$ and $b$, can be written as
\begin{equation}
 -{\mathcal L}_\nu^{\text{tree}} = h_{\alpha3}\bar\ell_\alpha^c{\rm i}\sigma_2\rho_3H + h_{\alpha0}\bar\ell_\alpha^c{\rm i}\sigma_2\rho_0H 
+ M_{\rho_3}{\rm Tr}\left(\bar\rho_3^c\rho_3\right) 
+ \frac{1}{2}M_{\rho_0}\bar\rho_0^c\rho_0 + {\rm H.c.},
\label{eq:tree:L}
\end{equation}
and interactions relevant for the one-loop contribution to the neutrino masses, shown in Fig.~\ref{fig:1:loop:nu:masses}~$c$, can be written as
\begin{equation}
 -{\mathcal L}_\nu^{\text{1-loop}} = -\frac{p_\alpha}{\sqrt{2}}\bar\ell_\alpha^c{\rm i}\sigma_2\text{Tr}(S_{8}\rho_8) + M_{\rho_8}{\rm Tr}\left(\bar\rho_8^c\rho_8\right) 
+ \frac{\lambda_S}{2} {\rm Tr}(S_{8}^\dag H)^2 + {\rm H.c.},
\label{eq:1:loop:L}
\end{equation}
where in \eqref{eq:tree:L} and \eqref{eq:1:loop:L} we have defined
\begin{eqnarray}
 h_{\alpha3} &=& c_\alpha \cos\theta - 3p_\alpha \sin\theta, \label{h3}\\
 h_{\alpha0} &=& \frac{\sqrt{15}}{2}\left( \frac{c_\alpha}{5}\cos\theta + p_\alpha \sin\theta \right). \label{h0}
\end{eqnarray}
Notice that in Eqs.~(\ref{h3}) and (\ref{h0}) we corrected the result of~\cite{0702287}.

The contribution of $\rho_{3}$ to the $T$ parameter is zero due to mass degeneracy of $\rho_{3}^{0}$ and $\rho_{3}^{\pm}$, because the contributions to neutral and charged Goldstones \cite{Barbieri:2006dq} cancel each other.

Only electrically neutral components of $\rho_i$ enter the mechanisms illustrated in Fig.~\ref{fig:1:loop:nu:masses}.
The quadratic coupling $\lambda_S$ arises from several terms in the Adjoint $SU(5)$ Lagrangian. Here we take it to be arbitrary, bounded only by perturbativity.

The resulting neutrino mass matrix is
\begin{eqnarray}
 M_{\alpha\beta}^\nu = v_0^2\left[ \frac{h_{\alpha3}h_{\beta3}}{M_{\rho_3}} 
+ \frac{h_{\alpha0}h_{\beta0}}{M_{\rho_0}}
+ \frac{p_{\alpha}p_{\beta}}{2M_{\rho_8}}\frac{\lambda_S}{4\pi^2}F\left(\frac{M_{S_8}}{M_{\rho_8}}\right) 
 \right]  \label{eq:nu_masses}
\end{eqnarray}
with
\begin{eqnarray}
 F(x)&=& \frac{x^2-1-\ln x^2}{(1-x^2)^2} \nonumber\\
 &=&-(1+2\ln x) + {\mathcal O}(x^2\ln x) \quad\quad {\rm for} \ \ x\ll1.
\end{eqnarray}
The first term in Eq.~(\ref{eq:nu_masses}) is generated in type III, the second term in I seesaw, as in~\cite{0702287}. 
The third term comes from one-loop coloured seesaw~\cite{0906.2950,1010.5802} 
in the limit of small mass splitting $|\lambda_S|v_0^2\ll (M_{8R}^2+M_{8I}^2)/2$, where $M_{8i}$ is the mass of $S_{8i}^0$ ($i=R,I$) ~\cite{precise_mass_formula}.

In~\cite{0702287,LG} the neutrino masses and LG were studied for $|\lambda_S|\ll1$ and hence the one-loop contribution to neutrino masses was neglected.
In this paper we are interested in the case of $\lambda_S\sim1$, in which the fermionic octet is relevant to the generation of neutrino masses.

Eqs.~\eqref{h3}--\eqref{eq:nu_masses} show that the number of 
light massive neutrinos 
in the considered model is two, 
which is equal to the maximal number of nonzero independent 
vectors in the plane defined by the vectors $c_\alpha$ and $p_\alpha$.\footnote{In addition, flavour universal values $c_\alpha = c$ and 
$p_\alpha = p$ for the couplings are not allowed, 
because in this case there would be only one massive neutrino.}
However none of $\rho_0$, $\rho_3$ and $\rho_8$ is decoupled from the generation of the neutrino masses due to its heavy mass or small couplings, as in the 
two right-handed neutrino models~\cite{twoN_R,Ibarra-Ross}. 
On the other hand, some of $\rho_i$ may not enter particular physical processes, e.g., $\rho_8$ decouples from LG, see section~\ref{section:LG}.
Such decoupling results in new physics effects, which are dependent on the Yukawa couplings 
$h_{\alpha i}$, in particular, the allowed parameter range of successful LG is enlarged, see section~\ref{sec:numerics}.

We remark that similar relation among the neutrino masses and LG may be derived in theory with two heavy Majorana neutrinos and one light sterile neutrino, 
which has specific mass and mixing terms with other neutrinos\footnote{In a toy model this sterile neutrino may have negative $Z_2$ parity and 
contribute to the neutrino masses due to the interaction with inert scalar doublet, see~\cite{precise_mass_formula}.}~\cite{progress}.

Using the linear relation
\begin{eqnarray}
 p_\alpha=a_3h_{\alpha3}+a_0h_{\alpha0}
\end{eqnarray}
with 
\begin{eqnarray}
 a_3=-\frac{1}{8\sin\theta}, \qquad  a_0=\frac{1}{4\sin\theta}\sqrt{\frac{5}{3}},
\end{eqnarray}
Eq.~(\ref{eq:nu_masses}) can be rewritten as
\begin{eqnarray}\label{eq:nu_masses2}
 M_{\alpha\beta}^\nu = v_0^2 h_\alpha \mathcal{M}_\rho^{-1}h_\beta^T ,
\end{eqnarray}
where $h_\alpha=(h_{\alpha3},h_{\alpha0})$ and 
\begin{eqnarray}
 \mathcal{M}_\rho^{-1} = \left(
\begin{array}{cc}
M_{\rho_3}^{-1}+a_3^2(M_{\rho_8}^{\text{eff}})^{-1} & a_3a_0(M_{\rho_8}^{\text{eff}})^{-1} \\
a_3a_0(M_{\rho_8}^{\text{eff}})^{-1} & M_{\rho_0}^{-1}+a_0^2(M_{\rho_8}^{\text{eff}})^{-1} \\
\end{array}
\right)
\end{eqnarray}
with
\begin{eqnarray}\label{M8eff}
 M_{\rho_8}^{\text{eff}} &=& 2\frac{4\pi^2}{\lambda_S}F^{-1}\left(\frac{M_{S_8}}{M_{\rho_8}}\right)M_{\rho_8} \nonumber\\
  &\simeq& \frac{4\pi^2}{\lambda_S} \left(\ln\frac{M_{\rho_8}}{M_{S_8}}-\frac{1}{2} \right)^{-1}M_{\rho_8} \qquad {\rm for}\ \ M_{S_8}\ll M_{\rho_8}.
\end{eqnarray}
In the limit $\lambda_S\to0$ matrix $\mathcal{M}_\rho^{-1}$ goes to $D_\rho^{-1}$ with $D_\rho=\text{diag}(M_{\rho3},M_{\rho0})$.

The neutrino mass matrix in Eq.~(\ref{eq:nu_masses2}) can be rewritten as
\begin{eqnarray}\label{eq:nu_masses_diag2}
 M_{\alpha\beta}^\nu = v_0^2\, h_\alpha^\prime D_\mathcal{M}^{-1}h_\beta^{\prime T},
\end{eqnarray}
by using the orthogonal transformation
\begin{eqnarray}
 \mathcal{M}_\rho^{-1} &=& Q^{T}D_\mathcal{M}^{-1}Q, \\
 h_\alpha &=& h^\prime_\alpha Q, \label{eq:h_hprime}
\end{eqnarray}
where $Q$ is a real orthogonal matrix and $D_\mathcal{M} = \text{diag}(M_{1},M_{2})$, where $M_{i}$ are the eigenvalues of $\mathcal{M}_\rho$.

\section{Neutrino masses from experiment}\label{section:nu_masses_exp}

The three neutrino mass spectra allowed by the oscillation data are
\begin{itemize}
\item Normal Hierarchical (NH)
\begin{eqnarray}
 m_1 \ll m_2<m_3, \quad m_2 = \sqrt{m_1^2+\Delta m_{\text{sol}}^2}, \quad m_3 = \sqrt{m_1^2+\Delta m_{\text{atm}}^2};
 \label{eq:NH}
\end{eqnarray}
\item Inverted Hierarchical (IH)
\begin{eqnarray}
 m_3 \ll m_1<m_2, \quad m_1 = \sqrt{m_3^2+\Delta m_{\text{atm}}^2-\Delta m_{\text{sol}}^2}, \quad m_2 = \sqrt{m_3^2+\Delta m_{\text{atm}}^2};
\end{eqnarray}
\item Degenerate, $m_i\simeq m_0\geq0.10$~eV ($i=1,2,3$);
\end{itemize}
where $\Delta m_{\text{sol}}^2 = 7.65\times10^{-5}$~eV$^2$ and $\Delta m_{\text{atm}}^2 = 2.40\times10^{-3}$~eV$^2$ 
are the mass-squared differences of solar and atmospheric neutrino oscillations~\cite{PDG2010}.

We use the standard Casas-Ibarra~\cite{Casas-Ibarra} parametrization of the Yukawa couplings $h^\prime_{\alpha i}$ as
\begin{eqnarray}\label{eq:hat_hd}
 h^\prime=v_0^{-1}UD_\nu^{1/2}\Omega D_\mathcal{M}^{1/2},
\end{eqnarray}
where $D_\nu={\rm diag}(m_1,m_2,m_3)$; the PMNS lepton mixing matrix $U$ can be written as
\begin{eqnarray}
 U = \left(
\begin{array}{ccc}
 c_{12}c_{13} & s_{12}c_{13} & s_{13}e^{-{\rm i}\delta} \\
 -s_{12}c_{23}-c_{12}s_{23}s_{13}e^{{\rm i}\delta} & c_{12}c_{23}-s_{12}s_{23}s_{13}e^{{\rm i}\delta} & s_{23}c_{13} \\
 s_{12}s_{23}-c_{12}c_{23}s_{13}e^{{\rm i}\delta} & -c_{12}s_{23}-s_{12}c_{23}s_{13}e^{{\rm i}\delta} & c_{23}c_{13}
\end{array}
\right)
 \times {\rm diag}(1,e^{{\rm i}\alpha_1/2},e^{{\rm i}\alpha_2/2}),
\end{eqnarray}
where $c_{ij}=\cos\theta_{ij}$, $s_{ij}=\sin\theta_{ij}$, with $\theta_{12}\simeq\pi/5.4$, $\theta_{23}\simeq\pi/4$ and $s_{13}^2<\pi/13$~\cite{PDG2010}, 
while the Dirac $\delta$ and Majorana $\alpha_j$ $CP$ violation phases are not fixed by the present experiments. 
Finally, the $3\times2$ complex matrix can be written as~\cite{Ibarra-Ross}
\begin{eqnarray}
 \Omega^{\text{NH}} = 
\left(
 \begin{array}{cc}
  0 & 0  \\
  \cos z & -\sin z \\
  \xi\sin z & \xi\cos z \\
 \end{array}
\right), \qquad
 \Omega^{\text{IH}} = 
\left(
 \begin{array}{cc}
  \cos z & -\sin z \\
  \xi\sin z & \xi\cos z \\
  0 & 0  \\
 \end{array}
\right)
\end{eqnarray}
in the normal and inverted hierarchy, respectively; $\xi=\pm1$.

\section{Leptogenesis}\label{section:LG}
\subsection{$CP$ asymmetry}

We define the $CP$ asymmetry as
\begin{equation}
 \epsilon_{\rho_3,\alpha} = \frac{\Gamma(\rho_3\to\ell_\alpha H^\dag)-\Gamma(\rho_3\to\bar\ell_\alpha H)}{\sum_\beta
\left[\Gamma(\rho_3\to\ell_\beta H^\dag)+\Gamma(\rho_3\to\bar\ell_\beta H)\right]}.
\end{equation}
In the considered model it is given by~\cite{Fukugita_Yanagida,StandardLG}
\begin{eqnarray}\label{eq:epsilon}
 \epsilon_{\rho_3,\alpha} = \frac{1}{8\pi\sum_\beta|h_{\beta3}|^2}{\rm Im}
\left[ h_{\alpha3}^*h_{\alpha0}\sum_\gamma h_{\gamma3}^*h_{\gamma0} \right]\, f(M_{\rho_0}^2/M_{\rho_3}^2),
\end{eqnarray}
where
\begin{eqnarray}
 f(x)= \sqrt{x}\left[ 1-(1+x)\ln\left(\frac{1+x}{x}\right)\right] 
= -\frac{1}{2\sqrt{x}} + \mathcal{O}(x^{-3/2}) \quad \text{for} \quad x\gg1
\end{eqnarray}
since the only non-vanishing contribution comes from the vertex correction~\cite{LG}.
We assume that the couplings of $|H|^2|S_8|^2$ and $|H^\dag S_8|^2$ interactions are small comparing to $\lambda_S$, 
so that two-loop effects in the $CP$ asymmetry can be neglected.

Using the total decay rate
\begin{eqnarray}
 \Gamma_{\text{D}} = \Gamma+\bar\Gamma = \sum_{\alpha,s}\Gamma(\rho_3^s\to\ell_\alpha H^\dag,\, \bar\ell_\alpha H),
\end{eqnarray}
where $s=-,0,+$ (signs of the components of $\rho_3$), the decay parameter $K$, given by the ratio of the decay width of a single $\rho_3$ component 
to the expansion rate when $\rho_3$ starts to become non-relativistic 
at $T=M_{\rho_3}$, can be written as
\begin{eqnarray}
 K= \frac{\Gamma_{\text{D}}}{3H|_{T=M_{\rho_3}}}, 
\end{eqnarray}
and rewritten as $K=\tilde m/m_*$ in terms of 
the rescaled decay rate~\cite{Davidson_Nardi_Nir}
\begin{eqnarray}\label{eq:tilde_m}
 \tilde m \equiv 8\pi\frac{v_0^2}{M_{\rho_3}^2}\frac{\Gamma_{\text{D}}}{3} = \frac{v_0^2}{M_{\rho_3}}\sum_\alpha|h_{\alpha3}|^2
\end{eqnarray}
and the rescaled Hubble expansion rate (equilibrium $\rho_3$ mass)
\begin{eqnarray}
 m_* \equiv 8\pi\frac{v_0^2}{M_{\rho_3}^2}H|_{T=M_{\rho_3}} \simeq1.08\times10^{-3}~\text{eV}.
\end{eqnarray}
For NH (IH) the strong washout regime requires
\begin{eqnarray}
 K\geq K_{\text{sol}\,(\text{atm})}\equiv m_{2(1)}/m_*\simeq8.1\ (46)\gg1.
\end{eqnarray}

Using Eqs.~\eqref{eq:h_hprime}, \eqref{eq:hat_hd}, \eqref{eq:epsilon} and \eqref{eq:tilde_m}, we have the rescaled decay rate
\begin{eqnarray}\label{eq:tilde_m_N}
 \tilde m &=& \frac{1}{M_{\rho_3}} \left[ Q_{11}^2M_{1}(m_a|c|^2 + m_b|s|^2) + Q_{21}^2M_{2}(m_a|s|^2 + m_b|c|^2)\right. \nonumber\\
  &+& \left.2Q_{11}Q_{21}\sqrt{M_{1}M_{2}}(m_b-m_a){\rm Re}(s^*c) \right],
\end{eqnarray}
and the total $CP$ asymmetry
\begin{eqnarray}\label{eq:epsilon3_N}
 \epsilon_{\rho_3} &=& \sum_\alpha\epsilon_{\rho_3,\alpha} \simeq \frac{(m_a+m_b)}{16\pi \tilde mv_0^2}\frac{\text{det}Q}{M_{\rho_0}} 
\left\{ (Q_{11}Q_{22}+Q_{21}Q_{12})M_{1}M_{2}(m_b-m_a){\rm Im}(s^2) \right.   \nonumber\\
&-& \left. 2\left[ Q_{11}Q_{12}M_{1}\left( m_a|c|^2+m_b|s|^2 \right) + Q_{21}Q_{22}M_{2}\left( m_a|s|^2+m_b|c|^2 \right) \right] 
\sqrt{M_{1}M_{2}}\, {\rm Im}(s^*c)\right\}  \nonumber\\
\end{eqnarray}
with $c=\cos z$, $s=\sin z$, and $a=2\,(1)$, $b=3\,(2)$ for NH (IH) neutrinos.

In the limit $\lambda_S\to0$ we get the standard expressions
\begin{eqnarray}\label{eq:tilde_m_0}
 \tilde m &\to& m_a|c|^2 + m_b|s|^2, \\
 \epsilon_{\rho_3} &\to& \frac{1}{16\pi \tilde mv_0^2} M_{\rho_3} (m_b^2-m_a^2)
\, {\rm Im}(s^2),  \label{eq:epsilon3_0}
\end{eqnarray}
where in Eq.~\eqref{eq:epsilon3_0} we corrected the result given in~\cite{LG} by the factor of 1/2.


\subsection{Boltzmann equations}\label{section:BE}

Boltzmann equations in the unflavoured regime ($M_{\rho_3}\gtrsim5\times10^{11}$~GeV) can be written as (for more details see~\cite{LG} and Refs. therein)
\begin{eqnarray}\label{eq:boltzmann_rho3}
 \frac{dN_{\rho_3}}{dz} &=& -(D+S)(N_{\rho_3}-N_{\rho_3}^{\rm eq})-2S_g(N_{\rho_3}^2-(N_{\rho_3}^{\rm eq})^2),\\
 \frac{dN_{B-L}}{dz} &=& -\epsilon_{\rho_3}D(N_{\rho_3}-N_{\rho_3}^{\rm eq})-WN_{B-L},
 \label{eq:boltzmann}
\end{eqnarray}
where $z=M_{\rho_3}/T$, and $N_X$ ($X=\rho_3,\,B-L$) is the number density of $X$ 
calculated in a co-moving volume containing one $\rho_3$ (all of its three components) in ultrarelativistic thermal equilibrium: 
$N_{\rho_3}^{\text{eq}}(T \gg M_{\rho_3}) = 1$. Initially, $N_{B-L}^{\text{eq}}(T \gg M_{\rho_3}) = 0$.

The abundance at equilibrium is given by
\begin{equation}
  N_{\rho_3}^{\text{eq}} = \frac{1}{2} z^2 \mathcal{K}_2 (z),
\end{equation}
where $\mathcal{K}_2$ is a modified Bessel function.\footnote{$\mathcal{K}_n (z)$ corresponds to the \texttt{BesselK[n, z]} function in Mathematica.}

The decay factor is
\begin{equation}
  D \equiv \frac{\Gamma_{\text{D}}}{H z} = 3 K z  \left\langle \frac{1}{\gamma} \right\rangle,
\end{equation}
where 
$H$ is the Hubble rate, and $\langle 1/\gamma\rangle$ is the thermally averaged dilation factor. The sum $D+S$, where 
$S$ is the contribution from Higgs-mediated scatterings, is given by
\begin{equation}
  D + S \simeq 0.3 K \left[ 1 + \left( \frac{M_{\rho_3}}{M_h} \right) z^2 \ln \left( 1 + \frac{a}{z} \right)  \right],
\end{equation}
where $M_h$ is Higgs mass and
\begin{equation}
  a = \frac{8 \pi^2}{9 \ln (M_{\rho_3}/M_h)}.
\end{equation}

For the gauge scattering of $\rho_3$~\cite{Strumia} we use the fit~\cite{LG}
\begin{equation}\label{eq:Sg}
  S_g\simeq 10^{-3}\frac{M_{\text{Pl}}}{M_{\rho_3}}\frac{\sqrt{1+\pi z^{-0.3}/2}}{(15/8+z)^2(1+\pi z/2)}e^{0.3z},
\end{equation}
where $M_{\text{Pl}}=1.221\times10^{19}$~GeV is the Planck mass.

The washout can be written as
\begin{equation}
  W(z) = j(z) \frac{3}{4} K \mathcal{K}_1 (z) z^3 + \Delta W (z),
\end{equation}
where
\begin{equation}
  j(z) = 0.1 \left( 1 + \frac{15}{8 z} \right) 
  \left[z \ln \left( \frac{M_{\rho_3}}{M_h} \right) \ln \left( 1 + \frac{a}{z} \right) 
  + \frac{1}{z} \right].
\end{equation}
The contribution of the non-resonant $\Delta L = 2$ processes to the washout is
\begin{equation}
  \Delta W (z) \approx 3 \times 10^{-3} \frac{0.186}{z^2} \left( \frac{M_{\rho_3}}{10^{10}~\text{GeV}} \right) \left( \frac{\bar{m}^2}{\text{eV}^2} \right),
\end{equation}
where $\bar{m}^2 \equiv m_1^2 + m_2^2 + m_3^2 = 2.5 \, (4.7) \times 10^{-3} $~eV for NH (IH) neutrinos.

After solving the Boltzmann equations \eqref{eq:boltzmann}-\eqref{eq:boltzmann_rho3}, we obtain $N_{B-L}^{\text{f}} = N_{B-L} (z \to \infty)$ 
(in our calculation we use final $z=10$, where the fit in Eq.~\eqref{eq:Sg} still applicable~\cite{LG}),
included in the final baryon asymmetry
\begin{equation}\label{eq:BEsolution}
  \eta_B = 3 \times \frac{3}{4} \times \frac{86}{2387} \times \frac{12}{37}\, N_{B-L}^{\text{f}} \simeq 3\times 0.88\times 10^{-2} N_{B-L}^{\text{f}},
\end{equation}
where 3 is the number of $\rho_3$ components, the dilution factor 86/2387 is calculated assuming standard photon production from the onset of LG till recombination~\cite{LG4pedestrians}, 
and 12/37 is the fraction of $B-L$ asymmetry converted into a $B$ asymmetry assuming that sphalerons remain in equilibrium until after the electroweak phase transition~\cite{Davidson_Nardi_Nir,Harvey_Turner}.

The result in Eq.~\eqref{eq:BEsolution} should be compared with the allowed values
\begin{equation}\label{eq:B-L_exp}
  5.1\times10^{-10}<\eta_B^{\text{BBN}}<6.5\times10^{-10}, 
\end{equation}
which come from the nucleosynthesis predictions and observed abundances of light elements~\cite{PDG2010}.

\section{Numerical Results}
\label{sec:numerics}

\begin{center}
 \begin{figure}[tb]
 \centering
\includegraphics[width=0.46\textwidth]{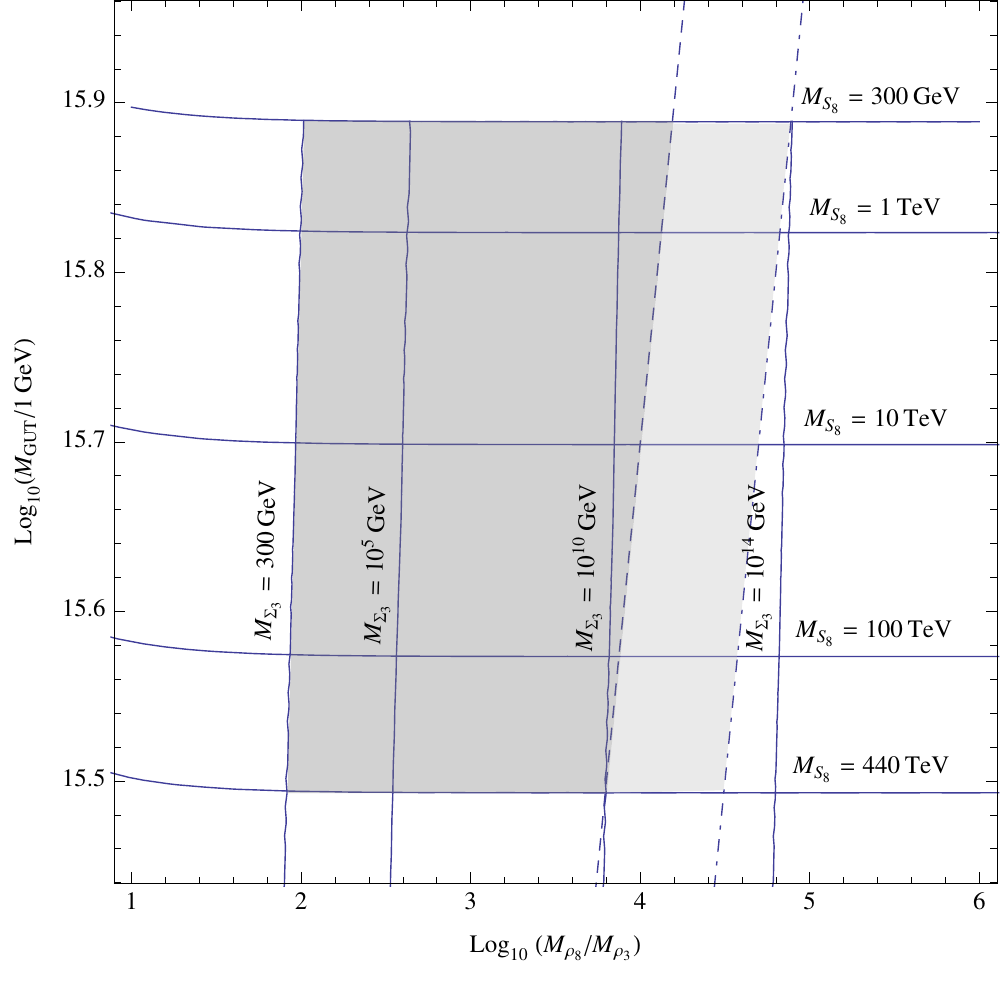}
   \caption{Parameter space allowed by unification at one-loop level for $M_{S_{(3,3)}}=10^{12}$~GeV. Dashed line shows the unflavoured and dot-dashed line the flavoured LG bound.}
   \label{fg:unification}
 \end{figure}
\end{center}
In the considered model the unification scale and the masses of $S_{(3,3)}$ and $S_8$ are constrained by the proton decay: 
$3\times10^{15}~\text{GeV}<M_{\text{GUT}}<8\times10^{15}$~GeV, $M_{S_{(3,3)}}>10^{12}$~GeV and $M_{S_8}<4.4\times10^{5}$~GeV~\cite{0803.4156}.
The parameter space allowed by the unification at one-loop level for $M_{S_{(3,3)}}=10^{12}$~GeV is shown in Fig.~\ref{fg:unification}, 
where the dashed line shows the bound from Eq.~\eqref{mass_relations} and the unflavoured LG constraint of 
$M_{\rho_3}\gtrsim5\times10^{11}$~GeV, and the dot-dashed line shows the flavoured LG bound of $M_{\rho_3}\gtrsim10^{11}$~GeV. 

\begin{center}
 \begin{figure}[tb]
 \centering
\includegraphics[width=0.36\textwidth]{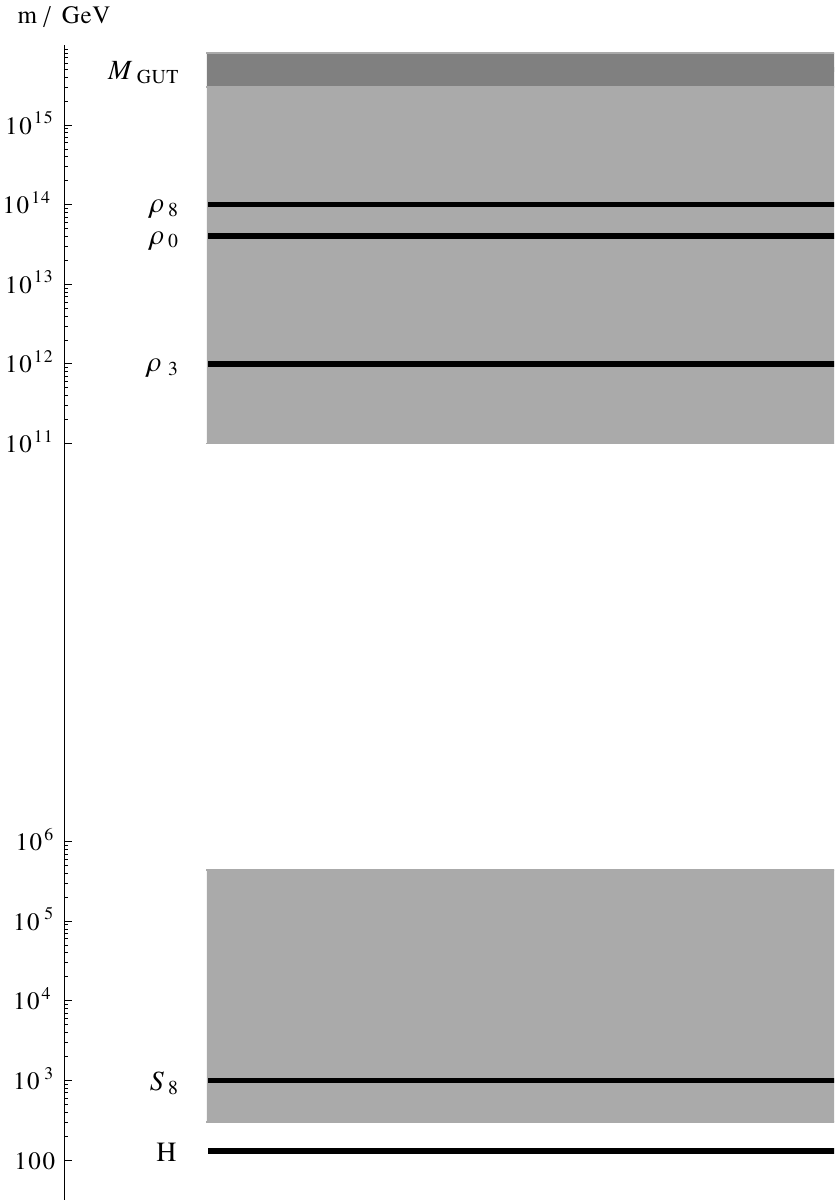}
   \caption{Mass spectrum of the considered model.}
   \label{fg:spectrum}
 \end{figure}
\end{center}
The mass spectrum of fermions and scalars in Renormalizable Adjoint $SU(5)$ is shown in Fig.~\ref{fg:spectrum}, 
where the lower light gray belt shows the allowed $S_8$ mass range from unification and proton decay constraints, and the upper light gray belt shows the allowed range for $\rho_i$ masses from LG. 
The dark gray belt shows the allowed range for the scale of unification. 
All the SM particle masses lie near or below $H$ mass $M_h$, while the rest of new particle masses in the theory can be set at the GUT scale, 
besides the mass of $\Sigma_3$ field, which favours unification and unflavoured (flavoured) LG within the wide region 
$300~\text{GeV}\lesssim M_{\Sigma_3}\lesssim1.5\times10^{11}(10^{14})$~GeV, where the lower bound comes from the collider searches.
The black lines for $S_8$ and $\rho_i$ in Fig.~\ref{fg:spectrum} show characteristic values of their masses.

\begin{center}
 \begin{figure}[tb]
 \centering
\includegraphics[width=0.48\textwidth]{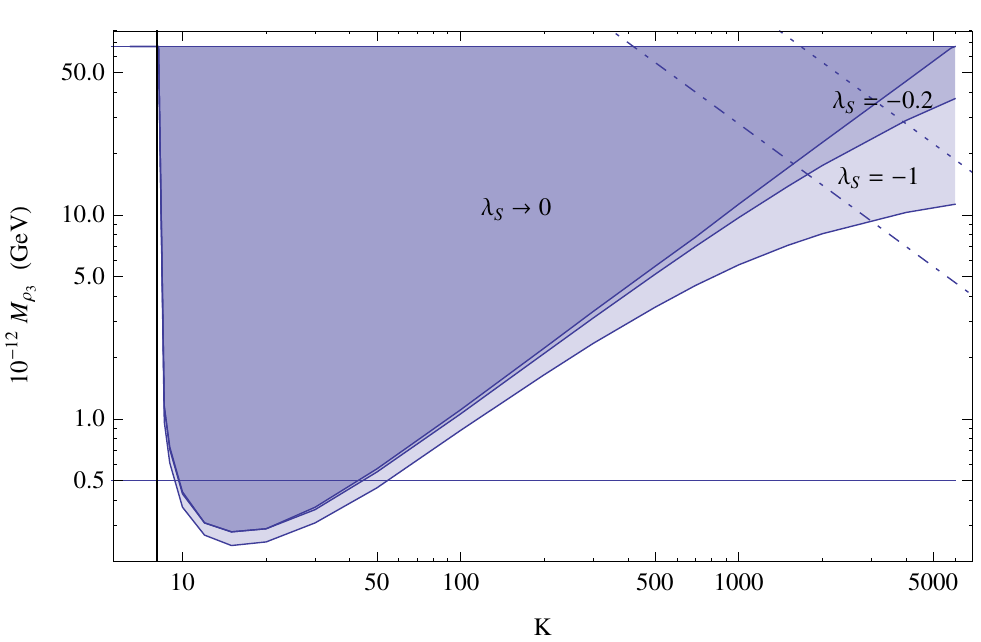}
\includegraphics[width=0.48\textwidth]{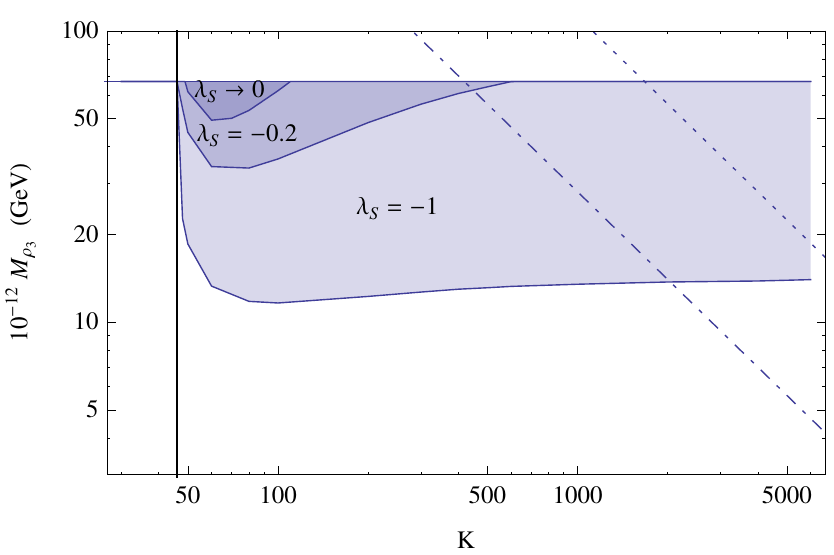}
   \caption{Left (Right): Allowed ranges for $M_{\rho_3}$ vs. $K$ in the case of NH (IH). 
See text for details.}
   \label{fg:BE}
 \end{figure}
\end{center}
Fig.~\ref{fg:BE} shows the allowed ranges for $M_{\rho_3}$ versus $K$ for the particular parameter sets of $\hat m=10^2$, $v_{45}=20$~GeV, and 
$\lambda_S=-1$, $-0.2$ and $|\lambda_S|\ll1$ (which reproduces the standard case in Eqs.~\eqref{eq:tilde_m_0}--\eqref{eq:epsilon3_0}), 
where the lighter region corresponds to larger $|\lambda_S|$.
The results are weakly sensitive to the Higgs mass within its expected range, and to $M_{S_8}$ due to only logarithmic dependence in Eq.~\eqref{M8eff}.
We use $M_h=180$~GeV and $M_{S_8}=10^{3}$~GeV. The upper bound for $\rho_3$ mass of $M_{\rho_3}\lesssim6.7\times10^{15}/\hat m$~GeV 
is derived using Eq.~\eqref{mass_relations} and Fig.~\ref{fg:unification}.
To get the lower bounds on $M_{\rho_3}$ for every given value of $K$, 
we maximize the solution~\eqref{eq:BEsolution} of the Boltzmann equations~\eqref{eq:boltzmann_rho3}--\eqref{eq:boltzmann} with the $CP$ asymmetry from Eq.~\eqref{eq:epsilon3_N} over the complex angle $z$, and find the minimal value of $M_{\rho_3}$ 
required to produce the observed baryon asymmetry in Eq.~\eqref{eq:B-L_exp} (its lower bound). 
The resulting curves in Fig.~\ref{fg:BE} are smoothened and given to few percent accuracy. 
The allowed region for LG in the case of vanishing loop contribution to the neutrino masses is shown by the darkest regions, 
which agrees with the result of~\cite{LG}. The lighter areas show the new allowed regions for LG for the larger values 
of $|\lambda_S|$ up to 1, when the loop contribution to the neutrino masses is significant.
The dashed lines show the lower bounds on $M_{\rho_3}$ for the parameter set $\hat m=10^2$, $v_{45}=30$~GeV and $\lambda_S=-1$.
The vertical lines show the lower bounds on $K$ in the strong washout regime. 
The dot-dashed and dotted curves show limits $M_{\rho_3}\leq v_0^2/(m_*K)$ and $M_{\rho_3}\leq 4v_0^2/(m_*K)$, which are derived from Eq.~\eqref{eq:tilde_m} 
using $\sum_\alpha |h_{\alpha3}|^2\leq1$ and 4, respectively.
Below these curves the couplings $h_{\alpha3}$ remain well perturbative.

We notice that the considered unflavored regime of LG is applicable for $M_{\rho_3}>5\times10^{11}$~GeV. Below this level, 
shown in Fig.~\ref{fg:BE} (left), the flavor effects should be taken into account.

We remark that the maximal allowed value of $M_{\text{GUT}}$, which is related to the mass of $S_8$, can be tested at the future proton decay experiments~\cite{0803.4156,0809.2106}.

\section{Summary and conclusions}
\label{sec:summary}

We have presented a detailed study of the neutrino masses and the baryogenesis via LG in Adjoint $SU(5)$, which is a well motivated model since testable by the proton decay experiments. $B-L$ asymmetry is generated by the decays of $SU(2)$ triplet $\rho_3$, 
with singlet $\rho_0$ running in the loop, while the neutrino masses are generated at the tree level 
by $\rho_3$ and $\rho_0$ exchanges and in one-loop coloured seesaw with propagating $SU(3)$ octets $\rho_8$ and $S_8$. For the values of the octet scalar $S_8$ to SM Higgs $H$ coupling $\lambda_S$ of order 1, 
the one-loop coloured seesaw is on the same order as the type I and III seesaw. This new contribution to the neutrino masses relaxes the Yukawa couplings 
for $\rho_3$ and $\rho_0$, and makes the allowed range for unflavored LG significant not only for the normal hierarchy of the neutrino masses, but also for the inverted one, 
in contrast to the case of small $\lambda_S$. 

The lightest neutrino, however, remains massless, because the coloured seesaw loop diagram is proportional to the Yukawa couplings of the type I and III seesaw.
For $|\lambda_S|\sim1$ the $CP$ asymmetry explicitly depends on several more parameters of Adjoint $SU(5)$, 
in comparing with the case of $|\lambda_S|\ll1$, such as: $\lambda_S$, 
the VEV $v_{45}$ of the representation ${\bf45}_H$, and the ratio $\hat m$ of the masses of adjoint fermions ${\rho_8}$ and ${\rho_3}$. This increases the predictive power and testability 
of the model.

We remark that some of the new particles, e.g., $\Sigma_3$ and $S_8$, 
may be light enough in the considered scenario to be tested at the LHC and the next generation of colliders~\cite{0809.2106,1010.5802}.

\appendix

\section{Field content of Adjoint SU(5)}
\label{app:fields}

Under $SU(3)_c\times SU(2)_L$, the scalar fields are decomposed as
\begin{align}
  {\bf 5}_H &= H_1 \oplus T_{(3,1)},\\
  {\bf24}_H &= \Sigma_8 \oplus \Sigma_3 \oplus \Sigma_{(3,2)} \oplus \Sigma_{(\bar 3,2)} \oplus \Sigma_{24}, \\
  {\bf45}_H &= S_8 \oplus S_{(\bar 6,1)} \oplus S_{(3,3)} \oplus S_{(\bar 3,2)} \oplus S_{(3,1)} \oplus S_{(\bar 3,1)} \oplus H_2.
\end{align}
and the matter fields are decomposed as
\begin{align}
  {\bf\bar5}_\alpha &= \ell_{\alpha L} \oplus (d_\alpha^c)_L, \\
  {\bf10}_\alpha &= (u_\alpha^c)_L \oplus q_L \oplus (e_\alpha^c)_L, \\
  {\bf 24} &= (\rho_8)_L \oplus (\rho_3)_L \oplus (\rho_{(3,2)})_L \oplus (\rho_{(\bar3,2)})_L \oplus (\rho_0)_L,
\end{align}
where $\alpha=1,2,3$ is the generation index, $\ell_L=(\nu_L, e_L)^T$ is the SM lepton doublet, $H_i$ are scalar $SU(2)$ doublets, 
and we denote $\rho_0\equiv\rho_{(1,1)}$, $S_8\equiv S_{(8,2)}$, $\Psi_3\equiv\Psi_{(1,3)}$ and $\Psi_8\equiv\Psi_{(8,1)}$, where $\Psi=\Sigma, \rho$. The VEVs are
\begin{align}
  \langle{\bf 5}_H\rangle &=  \frac{v_5}{\sqrt{2}} (0,0,0,0,1)^T, \\
  \langle {\bf 24}_H\rangle &= \frac{v_{24}}{\sqrt{30}}{\rm diag}(2,2,2,-3,-3), \\
  \langle(45_H)\rangle_j^{i5} &= \frac{v_{45}}{\sqrt{2}}\left[{\rm diag}(1,1,1,-3,0)\right]_j^i, \qquad \langle(45_H)\rangle_j^{in}=0,
\end{align}
where $i,j=1,\dots,5$, $n=1,\dots,4$, and relation $v=\sqrt{|v_5|^2+24|v_{45}|^2}=246$~GeV~\cite{Dorsner_VEV} provides proper masses of $SU(2)$ gauge bosons.

More details on the representations are given in the appendices A of \cite{0803.4156,0601023}. The scalar potential was studied in \cite{scalar_potential}, 
and the Yukawa couplings were analyzed in \cite{Dorsner_VEV,Yukawas}.

\section{Explicit form of used multiplets}\label{app:multiplets}
For the $SU(2)_L$ doublets and triplets we use the explicit forms
\begin{equation}
  \Psi = 
  \begin{pmatrix}
    \Psi^+ \\
    \Psi^0 \\
  \end{pmatrix}
   =
  \begin{pmatrix}
    \Psi^+ \\
    \frac{1}{\sqrt{2}}\left(\Psi_{R}^0+{\rm i}\Psi_{I}^0\right) \\
  \end{pmatrix}, \qquad
  \rho_3  = 
  \frac{1}{2}\boldsymbol{\sigma\cdot\rho}_3
  = \frac{1}{\sqrt{2}}
  \begin{pmatrix}
    \frac{1}{\sqrt{2}}\rho_3^0 & \rho_3^+ \\
    \rho_3^- & -\frac{1}{\sqrt{2}}\rho_3^0 \\
  \end{pmatrix},
\end{equation}
respectively, where $\Psi=H_1,\,H_2,\,S_8\dots$, $\rho_3^\pm=(\rho_3^1\mp {\rm i}\rho_3^2)/\sqrt{2}$, and $\sigma^{I}$ are the Pauli matrices.

The adjoint representation $24^i_j$ ($i,j=1,\dots,5$; $24^i_i=0$) can be written as
\begin{equation}
  {\bf 24} = 
  \frac{1}{\sqrt{2}}
  \begin{pmatrix}
    \frac{1}{\sqrt{2}}\boldsymbol{\lambda\cdot\rho}_8-\sqrt{\frac{2}{15}}\rho_0
    & \rho_{(3,2)} \\
    \rho_{(\bar3,2)} 
    & \frac{1}{\sqrt{2}}\boldsymbol{\sigma\cdot\rho}_3+\sqrt{\frac{3}{10}}\rho_0 \\
  \end{pmatrix},
\end{equation}
where $\lambda^{A}$ are the Gell-Mann matrices.

{\bf 45}$_H$ is defined by the conditions: $(45_H)_k^{ij}=-(45_H)_k^{ji}$ and $\sum_{i=1}^5(45_H)_i^{ij}=0$. 
We omit index $H$ of scalar representations in the following.
The explicit decomposition into SM multiplets can be deduced by considering the multiplet as the tensor product 
${\bf 45} = {\bf 5} \times {\bf 5} \times {\bf \bar{5}}$.
Below we will write the colour indices with letters from the beginning of the alphabet 
$a,b, \ldots=1,2,3$ and the weak indices as $r,s,\ldots=4,5$.
We have
\begin{align}
  { 45}^{ab}_{c} &= \epsilon^{abd} [S_{(\bar{6},1)}]_{dc} + \epsilon^{abd} [S_{(\bar{3},1)}]_{dc}, \\
  { 45}^{ab}_{r} &= \epsilon^{abc} [S_{(\bar{3},2)}]_{cr}, \\
  { 45}^{ar}_{b} &= \frac{1}{\sqrt{2}} S_{8}^{Ar} [\lambda^{A}]^{a}_{b} + H_{2}^{r} \delta^{a}_{b}, \\
  { 45}^{ar}_{s} &= \frac{1}{\sqrt{2}} S_{(3,3)}^{aI} [\sigma^{I}]^{r}_{s} + S_{(3,1)}^{a} \delta^{r}_{s}, \\
  { 45}^{rt}_{s} &= -\frac{3}{2} H_{2}^{t} \delta^{r}_{s},
\end{align}
where $[S_{(\bar{3},1)}]_{ab} = -[S_{(\bar{3},1)}]_{ba}$.




\end{fmffile}

\section*{{Acknowledgments}}

We thank Riccardo Barbieri, Anatoly Borisov, Konstantin Stepanyantz, Ilja Dor\u{s}ner, Enrico Bertuzzo and David Straub for useful suggestions and comments. 
This work was supported by the ESF Mobilitas 2 grant MJD140, Mobilitas 3 grant MTT8, ESF grants 8090 and 8943, and SF0690030s09 (KK); 
and by the EU ITN ``Unification in the LHC Era'',
contract PITN-GA-2009-237920 (UNILHC) and by MIUR under contract 2006022501 (DZ).

\end{document}